\documentclass[11pt, a4paper]{article}
\usepackage[utf8]{inputenc}
\usepackage[english]{babel}
\usepackage{soul}
\usepackage{csquotes,xpatch}
\usepackage[top=3cm, footskip=1cm, left=2cm, right=2cm]{geometry}
\usepackage{authblk}
\usepackage{amsfonts, amssymb, amsmath}
\usepackage{xfrac}
\usepackage{textcomp,marvosym}
\usepackage{nameref,hyperref}
\usepackage[switch]{lineno}
\usepackage{gensymb}
\usepackage{microtype}
	\DisableLigatures[f]{encoding = *, family = * }
\usepackage{mathtools}
\usepackage{booktabs} 
\usepackage{array}
\usepackage{float}
\usepackage{setspace}
\usepackage[hang, aboveskip=5pt,labelfont=bf,labelsep=period,justification=justified,textfont=small]{caption}
\usepackage{graphicx}
\usepackage[dvipsnames]{xcolor}
\usepackage{subcaption}
\usepackage{changepage}
\usepackage{physics}
\usepackage{soul}
\usepackage[numbers]{natbib}

\usepackage{lastpage,fancyhdr,graphicx}
	\fancyheadoffset[LR]{1.00in}
	\fancyfootoffset[LR]{1.00in}
\usepackage{epstopdf}
\newcommand{\EQ}{\begin{equation}}
\newcommand{\EE}{\end{equation}}
\newcommand{\EQA}{\begin{eqnarray}}
\newcommand{\EEA}{\end{eqnarray}}

\newcommand{\ant}{{\rm ag}}
\newcommand{\intr}{{\rm in}}

\newcommand{\syn}{{\rm syn}}
\newcommand{\ns}{{\rm nsyn}}
\newcommand{\vac}{{\rm vac}}
\newcommand{\comment}[1]{{\noindent #1}}

\newlength\savedwidth
\newcolumntype{+}{!{\vrule width 2pt}}

\title{Concepts and methods for predicting viral evolution} 

\author[a]{Matthijs Meijers\footnote{Equal contributions}}
\author[a]{Denis Ruchnewitz$^*$}
\author[a]{Jan Eberhardt}
\author[a]{Malancha Karmakar}
\author[b]{Marta~\L uksza$^\dagger$}
\author[a]{Michael~L{\"a}ssig\footnote{To whom correspondence should be addressed. Email: marta.luksza@mssm.edu, mlaessig@uni-koeln.de}}
\affil[a]{Institute for Biological Physics, University of Cologne, Z\"ulpicherstr. 77, 50937, K\"oln, Germany}
\affil[b]{Tisch Cancer Institute, Departments of Oncological Sciences and Genetics and Genomic Sciences, Icahn School of Medicine at Mount Sinai, New York, NY, USA}

\begin{document} 
%
\maketitle

\subsubsection*{Summary}

The seasonal human influenza virus undergoes rapid evolution, leading to significant changes in circulating viral strains from year to year. These changes are typically driven by adaptive mutations, particularly in the antigenic epitopes, the regions of the viral surface protein haemagglutinin targeted by human antibodies.
Here we describe a consistent set of methods for data-driven predictive analysis of viral evolution. Our pipeline integrates four types of data: (1) sequence data of viral isolates collected on a worldwide scale, (2) epidemiological data on incidences, (3) antigenic characterization of circulating viruses, and (4) intrinsic viral phenotypes. From the combined analysis of these data, we obtain estimates of relative fitness for circulating strains and predictions of clade frequencies for periods of up to one year. Furthermore, we obtain comparative estimates of protection against future viral populations for candidate vaccine strains, providing a basis for pre-emptive vaccine strain selection. Continuously updated predictions obtained from the prediction pipeline for influenza and SARS-CoV-2 are available on the website \href{https://previr.app}{previr.app}.

\bigskip
\noindent Keywords: antigenic evolution, population immunity, influenza vaccines, fitness models


\section{Introduction}

Global populations of the influenza virus encompass multiple antigenically distinct groups. The strains descending from a given mutant strain define a clade of the evolving viral population. A part of the genetic diversity encodes antigenic diversity: new strains escape human immunity generated by infection or vaccination with previous strains -- partially and into different directions. Antigenic diversity enables fast evolution driven by human immune pressure, leading to genetic and antigenic turnover on timescales of a few years. Thus, influenza evolution continuously decreases the protection of a given vaccine and requires regular vaccine updates. Given the time required for the development, production, and delivery of vaccines, current decisions to update the vaccine strain are made about nine months in advance of the next influenza season. Hence, optimal vaccines should be chosen in a pre-emptive way, to provide the best response to the strains circulating in the next season. These decisions benefit from modeling in two ways: to predict the antigenic characteristics of next year's circulating strains and to estimate the protection profile of different vaccine candidates against these strains. 
 
Two kinds of methods are currently used for influenza predictions in the context of vaccine strain selection. One strategy is to estimate growth differences between viral clades from a sequence- or phenotype-based fitness model. The original predictive fitness model for influenza uses amino acid mutations within antigenic sites of influenza hemagglutinin as markers of positive selection for antigenic escape, as well as mutations outside epitopes as a measure of negative selection (mutational load)~\cite{Luksza2014}. Antigenic assays measure the binding or neutralization capacity of host antisera against panels of test strains. More recently, suitably curated data from such assays~\cite{neher2016} have been integrated into antigenic fitness models for predictions~\cite{Morris2018,Huddleston2020, Meijers2023}. A complementary, model-free approach is to harvest genealogical trees built from viral sequences for information on the recent growth of genetic clades; this information can be extrapolated to predict near-future clade frequencies~\cite{neher2014predicting}.

Digesting viral data into successful predictions poses a fundamental problem: to assess the effects of genetic and phenotypic changes in viral strains facing the heterogenous population of all of our immune systems. Moreover, viruses and immune systems are coupled in a fast co-evolutionary process, so successful antigenic predictions have to track, and project into the future, the changes in viruses and in human population immunity~\cite{Meijers2023}. 

The computational analysis of influenza evolution, including predictions of near-future viral populations, is made possible by a unique combination of available data. First, sequence evolution of human influenza lineages is documented by worldwide surveillance over several decades~\cite{bao2008influenza}. Currently, about $40.000$ sequences of human influenza are sequenced each year. Second, influenza incidences, in part subtyped by lineage, is reported from multiple countries on a weekly basis. Third, the interactions of human and ferret antisera and viruses are extensively characterized by antigenic assays, specifically hemagglutination-inhibition (HI) and neutralization tests~\cite{WHOLabManual2010,WHOLabManual2011,pedersen2014hemagglutination,jorquera2019insights,cuevas2022vitro,Wang2022}. 

In this chapter, we describe an integrative suite of methods that combines all of these data to predict the prevalence and characteristics of emerging variants for fast-evolving respiratory viruses (Fig.~1). We focus on the analysis of human influenza, and we compare with parallel methods and results for SARS-CoV-2 where appropriate. We first describe the computational processing of viral sequence data, including the construction of genealogical trees and the evolutionary tracking of viral clades. As an application, we discuss how sequence and tree data can be combined to infer selection on the evolution of influenza proteins. Next, we describe the integration of epidemiological and antigenic data, which provide information on human immune pressure acting on viral evolution. We assemble this input into a data-driven fitness model for viral evolution and develop model-based computational estimates of vaccine protection. 

\begin{figure}[h!]
\centering
\includegraphics[width=.8 \linewidth]{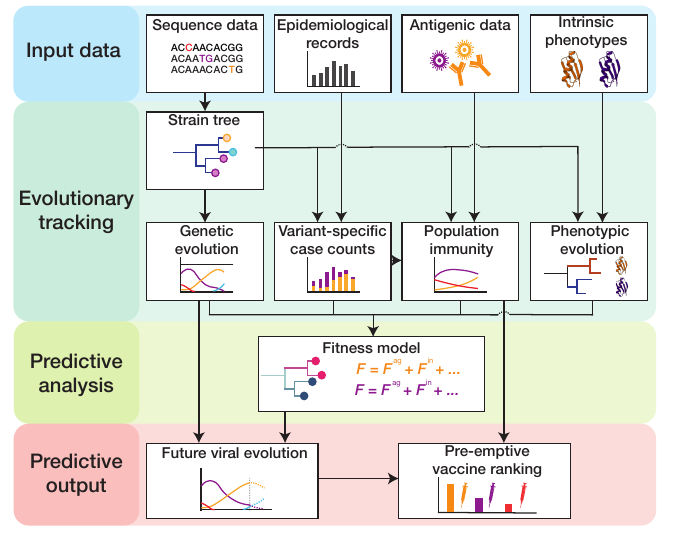}
\vspace*{0.3cm}
\caption{\textbf{Pipeline for predictive analysis of viral evolution}.
(1) {\bf Input data:} viral sequence data, epidemiological data, antigenic data, intrinsic viral phenotypes (protein stability, receptor binding, cell entry pathway, intra-cellular replication efficiency,~...). 
(2) {\bf Evolutionary tracking:} viral evolution (clade frequency trajectories and empirical clade fitness trajectories), clade-specific incidence dynamics, population immunity, and phenotypic evolution are tracked using a timed strain tree.  
(3) {\bf Predictive analysis:} fitness models are built from tracked immune data and tree-mapped sequence data and are calibrated by empirical fitness. 
(4) {\bf Predictive output:} fitness models serve to predict clade trajectories and antigenic evolution and to rank candidate vaccines by their predicted protection against future viral strains. 
}
\end{figure}

\section{Materials}
\subsection{Genetic data}
The repositories GenBank \cite{Benson2015} and GISAID \cite{Elbe2017} contain large numbers of influenza genome sequences (currently $> 400.000$ for human seasonal lineages). We use a primary curation procedure to exclude low-quality or incomplete sequences. We perform the following steps: 
(1)~Initial curation: raw, unaligned consensus nucleotide sequences with $> 1\%$ ambiguous characters or without a complete collection date are excluded.
(2)~Alignment: the remaining sequences are aligned with MAFFT \cite{Katoh2013}, using a reference isolate from GenBank (typically one of the oldest  isolates available, e.g., A/HongKong/1-5-MA21-1/1968 for influenza A(H3N2)).
(3)~Secondary curation: From the aligned set, we discard sequences containing $>1\%$ ambiguous characters.
(4)~Exclusion of outlier sequences (optional): We record pairwise sequence divergence and exclude isolates outside the 95th percentile within a given month. This step can filter out isolates from potential zoonotic events (transmissions from animals to human hosts). 
Together, these steps produce a set of aligned, high-quality sequences of viral isolates for all subsequent analyses (currently about $40.000$ sequences/year for human seasonal influenza).

\subsection{Epidemiological data}
The size of influenza epidemics varies considerably between years, lineages, and geographical regions~\cite{Caini2016,Dave2019,Zhou2019}. Factors determining the size of the influenza epidemic include the antigenic novelty of the predominant strains, as well as environmental, and socio-demographic factors \cite{coletti2018shifting}. We track the number of influenza cases using the epidemiological records provided by FluNet \cite{flunet2020}, a surveillance tool of the Global Influenza Surveillance and Response System (GISRS)  
FluNet reports weekly numbers of influenza virus infections that are detected at the country-level, totalling about $370.000$ influenza cases per year from $122$ different countries, regions, or territories. Detected influenza cases are tested for influenza type (A or B); subtyping into lineages is provided for a large fraction of the cases. For cases that are not subtyped, we assign the case data to the different influenza subtypes, using the distribution of positive tests at that time in the corresponding region. 

\subsection{Antigenic data}
Antigenic assays measure the neutralization capacity of antiserum from a human or animal host against a panel of test viruses. Common assays include the classical hemagglutination-inhibition (HI) assay \cite{pedersen2014hemagglutination} and various types of viral neutralization assays \cite{jorquera2019insights,cuevas2022vitro}. In an HI assay, binding of viral particles to red blood cells prevents their hemagglutination; antibody-antigen binding can disrupt hemagglutination. The assay is performed with serial 2-fold dilutions of the antiserum. The maximum $\log_2$ dilution titer at which the red blood cells do not agglutinate, $T$, provides a readout of the net binding between antiserum and virions~\cite{Spackman2020}. In neutralization assays, the antiserum is titrated in the presence of virions and target cells; the recorded titer measures the minimum serum concentration needed to inhibit cell infection by 50\%.
In both assays, higher titers correspond to better recognition of the test virus by the antiserum. 

In the common animal model for influenza, naive ferrets are infected with an influenza strain which we call the reference strain. Antigenic assays record the titer $T_i^j$ that quantifies the neutralization of virus $i$ by antiserum raised against reference strain $j$. Absolute titer values are confounded by non-antigenic effects, such as the potency of the antiserum or the avidity of the red blood cells used in the HI assay~\cite{bedford2014} . 
Therefore, we preprocess titer data to remove spurious variation in the homologous titers $T_j^j$: we apply shifts $T_i^j \rightarrow T_i^j + (\langle T_j^j \rangle - T_j^j)$, where $\langle T_j^j \rangle$ is the average homologous titer in the raw data (for influenza ferret assays, $\langle T_j^j \rangle \approx 11$). These standardized titers are used in the remainder of the text. The transformation leaves titer differences $\Delta T_i^j = T_j^j - T_i^j$, which measure evolutionary differences between viral strains, unchanged. For the predictive data analysis reported in this chapter, we use a published set of primary antigenic data produced by the Worldwide Influenza Centre, London~\cite{WIC}. This dataset contains nearly $30.000$ HI titers with test strains in the period  2010 -- 2023. Reference and test strains are carefully chosen to cover the antigenic diversity of the evolving population of circulating strains. 

Two further types of antigenic data have recently been included into evolutionary analysis: antigenic assays using polyclonal human antisera~\cite{welsh2023age,dadonaite2023full} and deep mutational scanning data (DMS)~\cite{welsh2023age,dadonaite2023full,cao2023imprinted}. In DMS experiments, a given backbone viral strain is used to construct a library of mutant strains, most of which contain a single point mutation away from the backbone strain. Then, in a high-throughput way, an escape score against a given antiserum is evaluated for each mutant \cite{starr2020deepa,starr2022deep}. In sections 3.3 and 3.4, we discuss the extension of predictive methods to these data and highlight challenges specific to each data type. 

\section{Predictive analysis of viral evolution}

\subsection{Tracking viral evolution}
The continuous, time-resolved surveillance of circulating influenza viruses is key to track evolutionary changes~\cite{Grubaugh2019} and is the basis for predictive analysis. Here we discuss a pipeline for evolutionary tracking that combines sequence data, epidemiological records, and phenotypic data. Based on a timed genealogical tree of the evolving viral population, we first construct trajectories of evolutionary change at the level of clade frequencies, clade fitness, and mutant allele frequencies. Second, integrating information from epidemiological records, we obtain trajectories of clade-specific case counts. Third, we discuss the mapping of molecular phenotypic data to the strain tree; the special case of antigenic phenotypes will be taken up in section 3.3.

\subsubsection*{Tree reconstruction and reassortment detection}
Genealogical trees are graphical representations of inferred evolutionary relationships among collected samples, illustrating their common ancestry and divergence over time \cite{Felsenstein2003}. 
Two types of tree reconstruction methods are most commonly used for viral data analysis: maximum parsimony approaches analyze sequences directly to infer a tree topology with the least evolutionary changes. Maximum likelihood methods use statistical models of sequence evolution to estimate the probability of a particular tree structure given the observed data, together with the probability of the evolutionary steps along the tree. Maximum likelihood approaches are generally favored for their accuracy and flexibility, despite their higher computational demands~\cite{Huelsenbeck1997}. 

We use a probabilistic, maximum-likelihood approach utilizing IQTree2 \cite{Minh2020}, which supports various evolutionary models. Following an inference of the tree topology, TreeTime \cite{Pavel2018} is employed to infer the time and sequences of internal nodes in the resulting trees. We perform the following steps: 
(1)~Determine a maximum-likelihood sequence evolution model (if unknown). ModelFinder \cite{Kalyaanamoorthy2017} identifies nucleotide substitution models that best fit the data based on likelihood metrics. For seasonal influenza, a suitable assumption is the general-time-reversible model (GTR) with invariant sites (I), nucleotide frequencies inferred from data (F), and site-specific rates of 8 categories (R8) (the corresponding model name for IQTree2 is GTR+F+I+R8).
(2) Construct a maximum-likelihood tree of curated genetic data using IQTree2. The output from step 2 is a nexus file that stores the inferred topology. 
(3) Use TreeTime~\cite{Pavel2018} to infer the ancestral sequences and the time of internal nodes of the genealogical tree.

The result of this procedure above is a timed, genealogical tree of available sequences (see Note~1 for further discussion). A strain tree for influenza A(H3N2) is shown in Figure 2A. The tree describes the tracked genetic evolution of the virus; individual nucleotide and amino acid changes are mapped on specific branches. Collected isolates are depicted on the tree as leaves (small circles). Internal nodes (positioned at  junctions of branches) represent the inferred common ancestors of the isolates. The tree is colored by clades, which are genetically distinct groups of strains with a recent common ancestor. In defining a set of clades, the appropriate level of coarse-graining is an intricate problem and depends on the goals of the analysis. Criteria for delineating clades include epitope mutations or measured antigenic differences, changes in protein stability or other phenotypes, or significant growth increase compared to the ancestral clade. All of these changes are possible markers of fitness differences relevant for evolutionary predictions (section 3.4). 

Influenza viruses have a segmented genome with multiple chromosome-like segments (8 segments for influenza A and B). These segments can mix during replication if a host is co-infected with two distinct strains. This process, called reassortment,  
significantly contributes to the genetic diversity of influenza, particularly for influenza type A viruses~\cite{Holmes2005, Nelson2006}.
Reassortment events break the assumption that each sequence originates from a single parent, which underlies tree reconstruction. Therefore, trees are commonly reconstructed for each segment independently; this step is followed by tree reconciliation analysis using visualization with tanglegrams, manual comparison~\cite{Holmes2005,Nelson2008}, or algorithmic approaches~\cite{Nagarajan2011,Yurovsky2011,Svinti2013,BarratCharlaix2022}. 
Here, we implement a joint procedure of tree reconstruction and reassortment detection for the genes encoding the surface proteins hemagglutinin (HA) and neuraminidase (NA), which accumulate more sequence changes than other genes. This procedure contains computationally efficient reassortment detection operating at the level of sequences: by comparison with earlier sequences, we identify isolates that have an exceptionally high number of changes in the NA segment, compared to a null model of sequence evolution by point mutations; a related approach has been proposed in ref.~\cite{Rabadan2008}. The algorithm partitions the sequence sample into bi-segment sequence clusters free of significant reassortments. For each cluster, we obtain a maximum-likelihood tree based on linked bi-segment sequences by the method described above, resulting in a subtree with a significantly reassorted sequence at its root. Finally, each cluster is mapped to a  parent cluster, using the maximum-likelihood HA ancestor of its reassorted root sequence. The output is a bi-segment tree with point mutations and significant reassortments mapped on its branches.

\subsubsection*{Viral frequency trajectories} 
A partitioning of the strain tree into clades is defined by a set of internal nodes, each representing the last common ancestor of a given clade. Each strain is mapped to exactly one clade, given by the closest clade-defining node in its ancestral lineage. The evolutionary success of a given clade, $\alpha$, is described by its time-dependent population frequency trajectory, $x^\alpha (t)$. We compute these frequency trajectories as follows. For each isolate $i$, we define a frequency $x_i (t) = w (t - t_i)/\sum_j w (t - t_j)$, where $t_i$ is the reported collection date and $w (\tau) = \exp (- \tau^4/4\sigma^2)$ with $\sigma = 45$d. The weight function $w(\tau)$ is a smooth sliding window and  a heuristic measure of the temporal distribution of cases related to each reported isolate. The frequency of each clade is then the sum of the frequencies of its strains, 
\EQ
x^\alpha(t) = \sum_{i \in \alpha}x_i(t). 
\label{eq:uncor_exclfreq}
\EE
Clade population frequencies, by definition, sum up to one, $\sum_\alpha x^\alpha(t) = 1$ for all $t$. 

For each clade $\alpha$, we can define the sublineage $S(\alpha)$ as the union of clade $\alpha$ and all its nested descendant clades. Sublineages are in one-to-one correspondence with subtrees of the strain tree. They define a second set of frequency trajectories, 
\EQ
X^\alpha(t) = \sum_{i \in S(\alpha)}x_i(t).
\label{eq:uncor_inclfreq}
\EE
In a fast-evolving system like influenza, successful clades will rapidly produce new sequence variation on their genetic background, which defines new, nested descendant clades. Therefore, we rarely observe clade frequencies $x^\alpha(t)$ close to 1. By contrast, a sublineage frequency $X^\alpha(t)$ includes all nested subclades; it reaches 1 when the sublineage has displaced all competing clades. Fig.~2B shows clade frequency trajectories $x^\alpha (t)$ for the set of clades marked on the strain tree as a stacked plot; Fig. 2C shows the corresponding sublineage frequencies $X^\alpha (t)$. 

A third set of trajectories tracks the time-dependent population frequency of a given (nucleotide or amino acid) allele $a$ at a given genome position $k$, 
\EQ
x_{k, a} (t) = \sum_{i | k,a} x_i(t),
\EE
where the sum runs over all strains carrying allele $a$ at position $k$. In contrast to clade and sublineage frequencies, allele frequencies count the occurrence of point mutations independently of the genetic background. Therefore, an observed increase of $x_{k, a} (t)$ may signal an allele-specific fitness advantage.

\begin{figure}[h!]
\centering
\includegraphics[width=.80\linewidth]{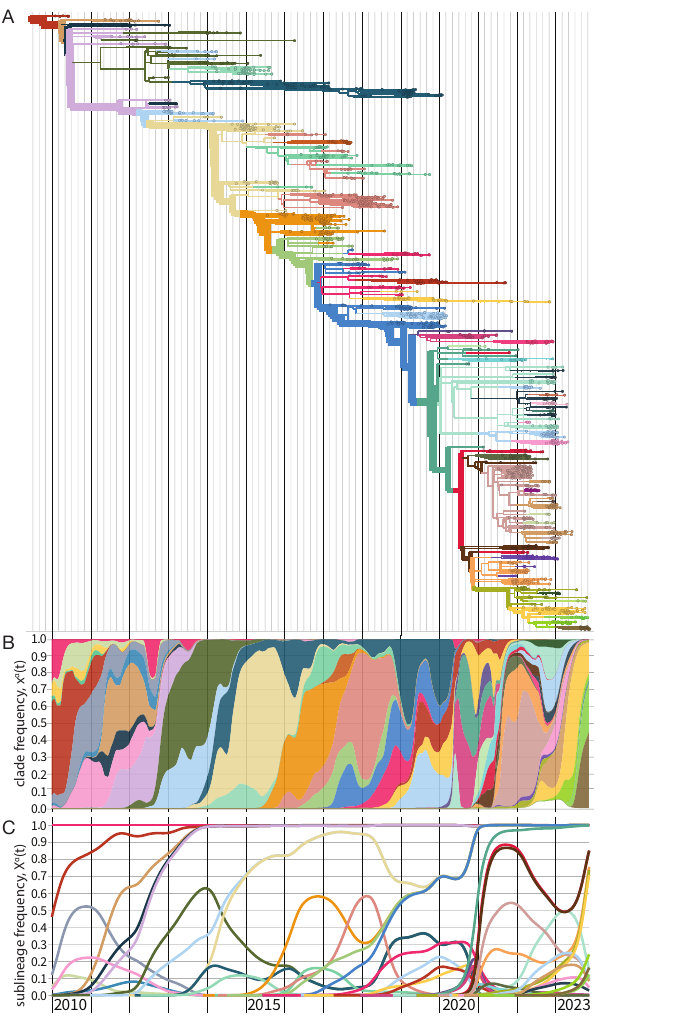}
\caption{{\bf Evolutionary tracking of influenza A(H3N2).} 
(A) Timed maximum-likelihood strain tree based on viral sequences from isolates with collection date from 2010 until end 2023. The tree is colored by clades $\alpha$ that reach a global frequency $X^\alpha (t) > 0.05$ and have potential antigenic differences to their parent clade (signalled by amino acid change(s) in an epitope or antigenic advance mapped from ferret antigenic data). 
(B) Stacked plot of global clade frequency trajectories, $x^\alpha (t)$, for this set of clades. 
(C) Global sublineage frequencies, $X^\alpha(t)$. 
}
\label{fig:tree}
\end{figure}

\clearpage

\subsubsection*{Empirical fitness trajectories} 
Given a set of clade frequency trajectories, $x^\alpha (t)$, we can also track the relative rates of change, 
\EQ
\hat f_\alpha(t)  = \frac{\dot x^\alpha (t)}{x^\alpha (t)}, 
\label{xdot2} 
\EE
where dots denote derivatives with respect to time. As discussed in section 3.4, the resulting trajectories $\hat f_\alpha (t)$ measure the relative fitness of each clade, relative to the mean population fitness, as a function of time. Clades with $\hat f_\alpha (t) > 0$ show a net increase in frequency, clades with $\hat f_\alpha (t) < 0$ a net decrease. A hat distinguishes these empirical fitness values derived from frequency tracking from the fitness models introduced below. The time-derivative is evaluated by numerical discretization around the time $t$ or by Bayesian inference from the timed isolate counts. A related measure of relative clade growth that can be evaluated from tree topology is the Local Branching Index (LBI) introduced in ref.~\cite{neher2014predicting}.

\subsubsection*{Regional frequency tracking}
Influenza strain populations show significant variation across regions, generating different sets of major regionally competing strains. This variation is of particular importance for emerging clades that originate in a specific region and have to be gauged in that context. Moreover, a regionally distributed population of circulating strains generates region-specific exposure histories that are relevant for the antigenic fitness model described below. 

Therefore, we track region-specific clade frequencies $x_r^\alpha(t)$, sublineage frequencies $X_r^\alpha(t)$, and allele frequencies $x_{r, k, a} (t)$; the frequencies in a given region $r$ are obtained by summation over all strains collected in that region. Major surveilled regions are roughly aligned with named WHO transmission zones; region-specific tracking can be refined to individual countries whenever this is relevant for a specific question. \comment{We note that the simple relation (\ref{xdot2}) between relative frequency change and empirical fitness is valid in a spatially homogeneous population. The regionally diverse seasonal epidemiology of human influenza viruses generates region- and time-dependent absolute growth rates of the viral population.} An inference procedure for empirical fitness trajectories $\hat f_\alpha (t)$ from region-specific data is discussed in section 3.4.

\subsubsection*{Incidence tracking}
By combining regional reported incidence numbers $I_r(t)$ obtained from FluNet~\cite{flunet2020} with regional clade frequencies $x_r^\alpha(t)$, we infer clade-specific incidence numbers, 
\EQ
I_r^{\alpha}(t) = I_r(t) \, x_r^\alpha(t), 
\label{inc} 
\EE
which enter the tracking of population immunity discussed in section 3.3. 

Regional incidence tracking can also serve to flag, and partially correct, differences in sequence sampling depth between regions. From sequence counts $N_r(t) = \sum_{i \, {\rm \in} \, r} w(t-t_i)$ and incidence data $I_r(t)$ in a set of regions $r$, we record weight factors $m_r(t) = I_r(t) / N_r(t)$ measuring the incidence per sequence count in each region. Given region-specific clade frequency trajectories, $x^\alpha_r (t)$, we can then compute incidence-weighed global clade frequencies, $x^\alpha (t) =  m^{-1} (t) \, {\sum_r m_r (t) \, x^\alpha_r (t)}$, where $m(t) = I (t) / N (t)$ is the ratio of global incidence and sequence counts. Incidence-weighed sublineage and allele frequencies are defined in an analogous way.  Alternatively, differences in sampling depth can also be inferred from local sequence diversity~\cite{Kathri2019,Smith2021}.

\subsubsection*{Tracking phenotypic evolution}
Several independently measurable molecular phenotypes of influenza proteins show evolutionary variation and may contribute fitness effects. Examples are protein stability, conformation of binding interfaces, as well as local solvent accessibility and glycosylation status of antigenic epitopes.  Using input from specific sequence data, recent computational methods serve to predict protein structures \cite{Alphafold2021,akdel2022structural}, the impact of mutations on protein stability and binding of protein complexes~\cite{Armita2024}, glycan shielding on the protein surface \cite{tsai2024glycan}, cross-neutralization~\cite{harvey2023bayesian}, variation of innate immune response~\cite{greenbaum2008,greenbaum2014}, as well as combinations of phenotypes relevant for growth~\cite{thadani2023learning}. Deep mutational scanning~\cite{starr2022deep, cao2023imprinted} and high-throughput in-vitro evolution~\cite{moulana2023genotype,zahradnik2021sars} provide experimental genotype-phenotype maps.

Tracking phenotypic evolution from these data requires mapping significant changes onto branches of the strain tree. Here we use a tree-guided interpolation procedure to obtain clade-specific phenotype values $E_\alpha$ from data $E_i$ measured for a limited set of strains. This procedure serves several purposes: (1) to average between measurements $E_i$ within clades with measured data, (2) to infer maximum-likelihood values $E_\alpha$ for clades without data, (3) to detect measurement outliers, and (4) to identify branches with significant phenotypic changes. A specific tracking algorithm is described in the context of antigenic phenotypes (section 3.3).


\subsection{Inference of selection on influenza proteins}
Sequence data organized in a genealogical tree and partitioned into clades with time-dependent sublineage frequencies contain a wealth of evolutionary information. In particular, they can serve to infer the speed of molecular evolution and the underlying selective forces, in different influenza genes and gene segments with specific functions. Second, they harbour information on the type of selection acting on viral clades that can guide the predictive fitness models to be developed below.

As an example of the inference method, Table~1 records the number of synonymous and non-synonymous nucleotide changes in the HA protein, mapped on a strain tree constructed from $>100.000$ isolates for influenza A(H3N2) in the period 1968-2020. The HA protein contains two domains, HA1 and HA2. The antigenic epitopes, which are the primary loci of interactions with the human immune system are found on the HA1 domain of the protein. These epitopes, designated A to E, contain a combined total of 62 amino acids \cite{wiley1981structural, bush1999predicting, munoz2005epitope, macken2001options}. Figure 3A shows the location of the five epitopes (A-E) on the monomeric protein structure of HA. Accordingly, we partition the HA point mutations into four sequence classes: antigenic epitopes A, B, D (located on the HA1 head), epitopes C, E (located closer to the HA1 stem), the receptor binding domain (RBD, which overlaps with epitopes A and B), and the remaining non-epitope sequence. Each mutation originates on a given branch of the tree, i.e., on a given genetic background, and defines a sublineage of descendant strains. We record the time-dependent frequency, $X(t)$, and record the number of all mutations that reach a given frequency $X$, denoted by $m(X)$. A mutation reaching frequency $X = 1$ (in practice, we use a threshold $X = 0.99$) signals fixation of the descendant sublineage; we denote the number of such fixations by $d$. In most cases, sublineage fixation also implies fixation of the defining allele change in the viral genome. Table~1 shows the numbers $m(0)$, $m(0.05)$, and $d$ for synonymous and non-synonymous changes in each sequence class. Here, the threshold frequency $X = 0.05$ serves to remove biases due to sampling depth (sampling in later years detects mutations at much lower initial frequency). Next, we compute the fraction of all mutations that reach a given frequency, $G(X) = m(X)/m(0)$. Throughout the influenza genome, most mutations remain at small frequency (i.e., close to the tips of the tree), only a small fraction reaches high frequency and eventual sublineage fixation (these mutations appear on the trunk of the tree). 

To quantify selection, we record the ratio 
\begin{equation} \label{ratio}
g (X) = \frac{G_\ns (X)}{G_\syn (X)}
\end{equation} 
as a function of the frequency $X$ for the different sequence classes. This measure, referred to as propagator ratio~\cite{strelkowa2012clonal}, gives the probability of non-synonymous mutations to reach sublineage frequency $X$, weighed by the corresponding probability for synonymous mutations. More generally, the ratio $g(X_2)/g(X_1)$ gives the conditional probability to reach frequency $X_2$, given an initial frequency $X_1$ already reached, again weighed by the corresponding probability for synonymous mutations. Hence, we obtain a frequency-sensitive measure of selection: the direction and amplitude of changes in $g(X)$ between frequencies $X_1$ and $X_2$ measure the amount of directional selection in the frequency interval $[X_1, X_2]$. The strongest signal of selection is contained in the propagator ratio evaluated over the full frequency interval $g = (d_\ns/m_\ns (0))/(d_\syn /m_\syn (0))$ (Table~1). We obtain a clear grading of selection in different HA gene segments of influenza A(H3N2): positive selection ($g > 1$) in antigenic epitopes (strongest in the head epitopes A, B, D, including the RBD sites overlapping with epitopes A and B), and negative selection ($g < 1$) in the remainder of the HA protein~\cite{strelkowa2012clonal}. Thus, selection analysis is able to identify the sequence loci subject to positive selection for immune escape evolution. Numerous other studies have identified positive selection in the antigenic epitopes of influenza \cite{fitch1991positive, bush1999positive, shih2007simultaneous, bhatt2011genomic, lin2012evolution, meyer2013cross, arunachalam2013detection, klingen2019structures} and in the RBD and nucleocapsid protein of SARS-CoV-2 \cite{rochman2021ongoing}. Similarly, purifying selection on viral protein evolution was mapped in a number of studies for influenza \cite{klingen2019structures, lin2019many, ghafari2022purifying} and SARS-CoV-2  \cite{ghafari2022purifying, neher2022contributions}.

\begin{table}[t]
\fontsize{10pt}{12pt}\selectfont
    \centering
  \caption{{\bf Selection inference for the HA protein of influenza A(H3N2).} From left to right: (1) $m_\syn(0)$, number of all synonymous mutations; (2) $m_\syn(0.05)$, number of synonymous mutations reaching frequency $X > 0.05$; (3) $d_\syn$, number of synonymous mutations reaching sublineage fixation; (4) $m_\ns(0)$, number of all non-synonymous mutations; (5) $m_\ns(0.05)$, all non-synonymous mutations reaching sublineage frequency $0.05$; (6) $d_\ns$, number of non-synonymous mutations reaching sublineage fixation; (7) propagator ratio, $g$; (8) mutation ratio, $q$. Data from influenza A(H3N2) strain tree 1968-2020, 120K isolates.}

  \label{tab:table1}
  \vspace{10pt} 
        \begin{tabular}{l||c|c|c|c|c|c||c|c}
		sequence class & \multicolumn{2}{c|}{$m_{\syn}$} & $d_{\syn}$ & \multicolumn{2}{c|}{$m_{\ns}$} & $d_{\ns}$ & $g$ & $q$
		\\
		& \multicolumn{1}{c|}{0.0} & \multicolumn{1}{c|}{0.05} & & \multicolumn{1}{c|}{0.0} & \multicolumn{1}{c|}{0.05} & & & 0.05 
		\\
		\hline
		\hline
		epitope A, B, D & 4048 & 146 & 17 & 6160 & 483 & 120 & 3.06 & 1.58
		\\
		\hline
		epitope C, E & 1669 & 68 & 8 & 3195 & 184 & 36 & 1.77 & 1.29
		\\
		\hline
		receptor binding domain & 1998 & 103 & 15 & 1766 & 140 & 27 & 2.40 & 0.65
		\\
		\hline
		non-epitope & 23499 & 1066 & 161 & 13982 & 542 & 40 & 0.45 & 0.24
		\\
		\end{tabular}
  \vspace{10pt} 
\end{table}

To obtain a more detailed picture of selection, we plot the propagator ratio $g(X)$ as a function of the sublineage frequency $X$. We observe that most of the selection signal is in the frequency range $X \lesssim 0.3$. For higher frequencies, the pattern of $g(X)$ flattens, indicating that initially successful escape variants can be driven to loss by the subsequent dynamics~\cite{strelkowa2012clonal}. Two factors contribute to this effect, the relative strength of which delineates modes of evolution under directional vs.~diversifying selection ~\cite{Frazao2022}. First, clonal interference -- the selective competition between co-existing clades -- introduces strong collective effects at high sublineage frequencies that override the selection coefficients of individual mutations and introduce so-called hitchhiking effects: neutral and even moderately deleterious mutations can reach high frequencies in a successful sublineage, while mutations under positively selected can be outcompeted by another clade~\cite{strelkowa2012clonal}. Second, antigenic selection introduces time-dependent, non-linear effects \cite{Luksza2014, Meijers2023}: immune waning and previous infections deplete the pool of susceptible human hosts available to each of the competing clades. In particular, the self-coupling of an initially successful clade (that is, the consumption of susceptibles by previous infections from that clade and closely related clades) generates a systematic fitness decline at later times~\cite{Meijers2023,barrat2024eco}.

\begin{figure}[t!]
\centering
\includegraphics[width=.8\linewidth]{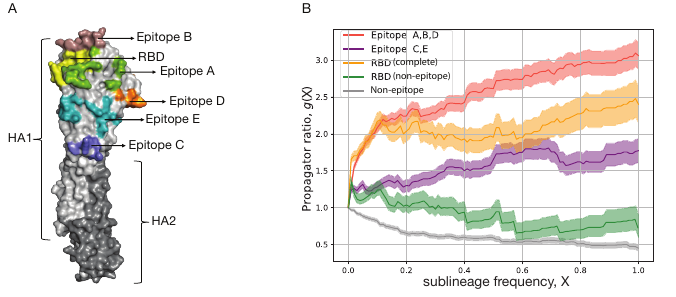}
\caption{\textbf{Selection on protein segments of influenza HA.}
(A) Protein structure of HA with marked regions: subdomains HA1 (light grey) and HA2 (dark grey), antigenic epitopes A -- E, receptor binding domain (RBD). 
(B) Frequency propagator ratio, $g(X)$, as a function of the sublineage frequency threshold $X$, for different HA sequence classes of influenza A(H3N2). Predominantly positive selection ($g > 1$) is inferred for the epitopes including overlapping RBD sites, overall negative selection ($g < 1$) for the non-epitope HA sequence. Shading indicates standard error margins.
}
\end{figure}

The propagator ratio $g$ is conceptually and computationally related to the McDonald \-- Kreitman test of selection \cite{mcdonald1991adaptive}. We note that probability ratios of the form (\ref{ratio}) are insensitive to uncertainties in entry frequency and timing of polymorphism histories, as well as to frequency-dependent bias in polymorphism numbers, as long as this bias does not depend on mutation class. 

Alternatively, we can directly compare the numbers of synonymous and non-synonymous changes above a given sublineage frequency threshold. We compute the ratio 
\EQ
q (X) = \frac{m_\ns (X)}{q_0 \, m_\syn (X)}, 
\EE 
where $q_0$ is the expected value under neutral genome evolution. We obtain $q_0 \approx 2.1$ by numerical evolution of influenza HA, using the reference sequence A/Hong Kong/1-5-MA21-1/1968 (see 2.1) as starting point, transition and transversion rates estimated by IQTree2, and discarding mutations that  produce a stop codon. Other computational implementations of neutral evolution are discussed in refs. \cite{yang2000estimating, goldman1994codon}. Evaluating $q(0.05)$ in different sequence classes yields again a signal of positive selection in the antigenic epitopes, albeit weaker than by the propagator ratio method, and of negative selection in the remainder of the HA sequence (Table~1). The $q$ measure is related to the classical dN/dS method \cite{miyata1980nucleotide, li1985new, nei1986simple}. Previous theoretical work has highlighted potential biases and inaccuracies in dN/dS estimates \cite{rocha2006comparisons,  Plotkin2008, mugal2014time, meyer2015time, rahman2021weak}. In a genealogical tree, the $q$ ratio measures originations, rather than substitutions as in a strain tree across different species. With a given threshold frequency $X$, it captures only the part of selection acting below $X$. 

The inference of selection can be refined to individual amino acid changes, integrating computational predictions of phenotypic effects~\cite{Armita2024, harvey2023bayesian, thadani2023learning} and experimental genotype-phenotype maps~\cite{starr2022deep, cao2023imprinted,moulana2023genotype,zahradnik2021sars}; see the discussion in section 3.1. A common challenge for experimental and computational analysis is to capture not only the effects of single mutations, but of combinations of multiple mutations that arise in circulating strains.

\subsection{Tracking population immunity}
The antigenic interaction between pathogen and host is a complex phenotype. The neutralization of a virus during an infection depends on the viral strain and on the individual immune system. The amount of neutralization shows considerable variation across the viral and the immune population. Moreover, both populations are highly dynamic: different viral clades and different classes of immune systems become prevalent over time. Here we develop a method to track these co-evolutionary dynamics, using a combination of molecular antigenic data, viral frequency data, and epidemiogical records. 

The tracking of antigenic evolution at the level of molecular interactions, here cross-neutralization titers, is jointly organized along a strain tree of primary infections, generating immune classes~\cite{Meijers2023}, and another strain tree of secondary infections. Next, we define two antigenic interaction measures at the population level: population immunity trajectories, again organized by immune class and viral clade, and population immunity profiles that summarize the net immune pressure on a given viral clade. We first develop the tracking method for antigenic data from ferrets; the application to human antigenic data is discussed at the end of the section. A glossary of all antigenicity measures used in this Chapter is given in Note~2.

\subsubsection*{Tracking antigenic evolution}
Antigenic titers $T_i^j$ from ferrets are obtained from immune-naive animals infected with a reference strain $j$, whose antisera are subsequently assayed against a panel of test strains $i$ representing potential secondary infections (see section 2.3). Titers from binding assays measure the reduced free energy of binding between virus and polyclonal antiserum; titers from neutralization assays remain strongly correlated with binding~\cite{veguilla2011sensitivity}. Reference strains and test strains are judiciously chosen from the evolving viral population but cover only a small fraction of the circulating strains. Here we describe a tree-guided interpolation method that produces a complete antigenic interaction matrix at the level of clades, $T^\kappa_\alpha$ (Fig.~4). Here $\kappa$ is the clade of the primary infecting strain $j$, in the following also called immune class, and $\alpha$ denotes the clade partitioning of test strains $i$. The completed cross-neutralization matrix also quantifies local antigenic evolution: the antigenic advance from clade $\alpha$ to clade $\beta$, 
\EQ
\Delta T^\kappa_{\alpha \beta} = T^\kappa_{\alpha} - T^\kappa_\beta,
\label{DeltaT} 
\EE
measures the immune escape effect of the mutation(s) separating these clades on the strain tree, as seen from immune class $\kappa$. Thus, antigenic advance is the change in the free energy of functional antibody-antigen binding induced by viral sequence evolution.

In a first step, we infer matrix elements $T^\kappa_\alpha$ for clade pairs $(\alpha, \kappa)$ containing measurements $T_i^j$. We use a quadratic optimization procedure with a cost function 
\EQ
S_1  =   \sum_{\alpha, \kappa} \sum_{(i,j) \in (\alpha, \kappa)} \left( T_i^j - \Delta_i - \Delta^j - T_\alpha^\kappa\right)^2  
  +   \frac{1}{\sigma_{\rm av}^2} \sum_i (\Delta_i)^2 + \frac{1}{\sigma_{\rm pot}^2} \sum_j (\Delta^j)^2. 
\EE
Here $T_i^j$ are standardized titer data, as described in section 2.3. The correction terms $\Delta_i$ and $\Delta_j$ (with the constraint  $\sum_i \Delta_i = \sum_j \Delta^j = 0$) account for strain-specific confounding factors caused by viral avidity and serum potency effects, respectively; the coefficients $\sigma_{\rm pot}$ and $\sigma_{\rm av}$ tune the amplitude of these corrections~\cite{bedford2014}. As shown in Fig.~4AB, this step achieves a partial data completion, setting titers $T^\kappa_\alpha$ close to the average of the measurements covering the corresponding clade pair $(\alpha, \kappa)$. In a second step, we infer the remaining $T^\kappa_\alpha$, which belong to clade pairs $(\alpha, \kappa)$ without measurements. We use an interpolation procedure that minimizes the cost function 
\EQ
S_2  =  \lambda_0   \sum_{\alpha,\kappa} \left( \Delta T^{\kappa}_{\mathcal{A}(\alpha) \alpha} \right)^2 
+ \lambda_1 \sum_{\alpha, \kappa} \left( \Delta T^{\mathcal{A}(\kappa)}_{\mathcal{A}(\alpha) \alpha} - \Delta T^{\kappa}_{\mathcal{A}(\alpha) \alpha} \right)^2 ,
\EE
where $\mathcal{A}(\alpha)$ is the parent of clade $\alpha$ and $\mathcal{A}(\kappa)$ is the parent of clade $\kappa$. This step produces a complete interaction matrix $T^\kappa_\alpha$ (Fig.~4C). We test the accuracy of the inference scheme using a $90/10$ training/test partitioning of the antigenic data and setting $\Delta_i = 0$ for strains that are not in the training set. We find a mean square error $\langle (T_i^j - \Delta_i -\Delta^j - T_\alpha^\kappa)^2 \rangle = 0.55 \pm 0.03$ (95\% confidence interval) in the test set. This should be compared with the expected measurement error of $\pm 1$ titer in an HI assay.  We conclude that the curation and completion scheme reduces experimental noise; that is, the inferred titer values $T_\alpha^\kappa$ are likely to be more informative than individual antigenic measurements $T_i^j$.  

The inference procedure presented here allows for antigenic advance between viral clades to depend on the immune class, $\Delta T^\kappa_{\alpha \beta} = T^\kappa_\alpha - T^\kappa_\beta$. It generalizes a previous method that infers antigenic advance with the constraint $\Delta T^\kappa_{\alpha \beta} = \Delta T_{\alpha \beta}$, where $\Delta T_{\alpha \beta}$ is a sum of uni-valued increments on the tree branches~\cite{neher2016}. The increased method complexity is warranted by the data: from the ferret antigenic data for influenza, we find that some clade pairs $(\alpha, \beta)$ show a significant antigenic advance $\Delta T^\kappa_{\alpha \beta}$ in some immune classes $\kappa$ but not in others; a similar effect for SARS-CoV-2 is discussed in ref.~\cite{Meijers2023}. 
 
\begin{figure}[t]
\centering
\includegraphics[width=.9\linewidth]{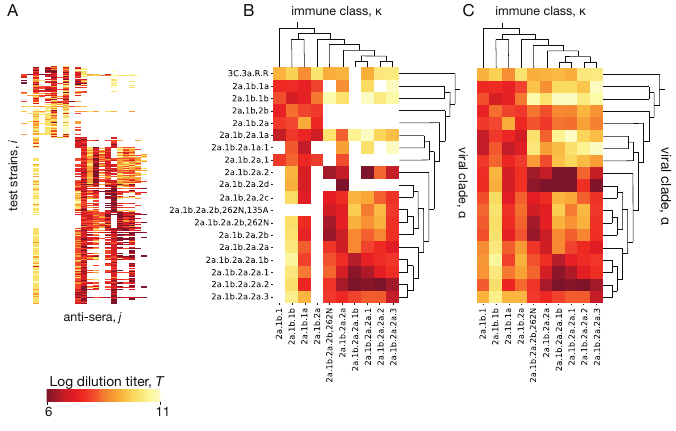}
\caption{\textbf{Inference of cross-neutralization}. (A) Set of raw titers $T_i^j$ between viral test strains $i$ and reference strains $j$ from antigenic assays. 
(B)~Coarse-grained matrix elements $T_\alpha^\kappa$ for clade pairs $(\alpha, \kappa)$ containing measured strain pairs $(i,j)$. 
(C)~Full inferred cross-neutralization matrix $T_\alpha^\kappa$ covering major clades in the period 2020--2024; missing values are inferred using a tree-guided interpolation scheme. Ferret antigenic data for influenza A(H3N2) are from ref.~\cite{WIC}, clade names follow WHO nomenclature. 
}
\end{figure}

\subsubsection*{Inference of cross-immunity}
How can the tracking of molecular antigenic evolution be used to build dynamical measures of antigenic interactions at the level of individuals and population segments? Here we relate clade-specific binding or neutralization titers, $T_\alpha^\kappa$ to the corresponding cross-immunity factors, $c_\alpha^\kappa$, with measure the relative reduction of susceptibility to infection by viral strains from clade $\alpha$, given a primary infection by a strain from clade $\kappa$ \cite{andreasen1997dynamics,gog2002dynamics,gog2002status}. We use a nonlinear map,
\EQ
c_\alpha^\kappa = H (T_\alpha^\kappa) = \frac{1}{1 + \exp[(T_{50} - T_\alpha^\kappa)/ \gamma]} , 
\label{H} 
\EE
with system-specific parameters $T_{50}$, $\gamma$ (for standardized titers from influenza ferret data, $T_{50} = 8$ and $\gamma = 1.4$). This form  follows the thermodynamic relation between binding probability and free energy, a rationale familiar from biophysical fitness models~\cite{Gerland2002,berg2004adaptive,zeldovich2007protein, Mustonen2008,manhart2015protein,chi2016selection,rodrigues2016biophysical,held2019survival}. 
However, equation (\ref{H}) says more than molecular binding kinetics: it relates a molecular variable, $T$, to an organismic effect, $c$ (the likelihood to get the flu). This relation is a heuristic; its validity has to be corroborated by data from both scales. For viral infections, a sigmoidal relationship between HI titers and measured protection from an infective challenge has first been reported in ref.~\cite{Hobson1972}; later studies confirmed this form in several viral systems~\cite{Coudeville2010, rotem2018evolution,Meijers2021,Khoury2021,Meijers2023}. 
In using ferret antigenic data as an input for cross-immunity factors of human individuals, we assume that the ferret titers from immune class $\kappa$ are a reasonable proxy for the neutralization capacity of humans who had their last infection by a strain from clade $\kappa$ (Note~3).

\subsubsection*{Population immunity trajectories and profiles}
To scale cross-immunity to the population level, we weigh the coefficients $c_\alpha^\kappa$ with epidemiological data of past infections. This defines a family of cross-immunity trajectories~\cite{Meijers2023}, 
 \EQ
 C_{\alpha}^\kappa(t) = c^\kappa_{\alpha} \, r_\kappa(t), 
\label{C_alpharho} 
 \EE 
which describe the population immunity against viral clade $\alpha$ generated by the population segment in immune class $\kappa$. The time-dependent factors $r_\kappa (t)$ determine the weight of immune class $\kappa$ in the human population at time $t$. We obtain these weights by summation over past infections, $r_\kappa (t) = \sum_r \int^t I^\kappa_r (t') \, K_\kappa (t - t') \, dt'$, using clade-resolved incidence numbers given by equation~(\ref{inc}). The kernel $K$ accounts for immunodominance-changing infections or vaccinations between $t'$ and $t$; similar kernels describe waning of immune protection for SARS-CoV-2~\cite{Meijers2023}.  

By summation over immune classes, we obtain the population immunity profile, 
 \EQ
 \bar C_{\alpha}(t) = \sum_\kappa c_\alpha^\kappa \, r_\kappa(t),
 \label{C_ferret} 
 \EE
which measures the total population immunity against clade $\alpha$ at time $t$. This profile describes the immune pressure on viral evolution and will be a key input to viral fitness models (section 3.4).

\subsubsection*{Tracking of human antigenic data} 
Antigenic data based on human antisera give a more direct representation of the human immune landscape, which is marked by complex histories of previous infections and vaccinations. Antigenic assays performed on a sample of antisera from individuals ($n = 1, \dots, N$) and a set of test viruses produce a set of titers $T^n_\alpha$. We then compute a set of cross-immunity profiles specific to each individual,
\EQ
c^n_\alpha = H (T^n_\alpha), 
\label{cnalpha} 
\EE
by applying a nonlinear map analogous to equation~(\ref{H}) with parameters $T_{50}, \gamma$ appropriate for human data~\cite{Hobson1972,Coudeville2010,Khoury2021,Meijers2023}. Given a sufficiently large and representative population sample collected around a given time $t$, 
the sample-averaged immunity profile, 
\EQ
 \bar C_{\alpha}(t) = \frac{1}{N} \sum_{n=1}^N c^n_\alpha , 
 \label{C_human}
 \EE
may provide a good approximation of the underlying population immunity profile. 

Recently, deep mutational scanning (DMS) experiments have been developed for the analysis of antigenic data from human population samples~\cite{welsh2023age,dadonaite2023full}. These high-throughput experiments generate antigenic escape mutations away from a backbone genome under immune pressure from an individual's antiserum and measure the fraction of unbound viral particles. From this readout, free energy effects
of individual mutations can be extracted~\cite{yu2022biophysical}, which correlate well with measured titer differences between the backbone strain and the single-mutant strain~\cite{welsh2023age}. Assuming additivity of local free energy changes, DMS data can be extrapolated to map the antigenic advance of strains with multiple mutations away from the backbone strain~\cite{yu2022biophysical}. Coarse-graining to the level of clades, we can track the advance $\Delta T^n_{\alpha \beta}$ from the backbone clade $\alpha$ to mutant clades $\beta$, as seen in a given individual $n$, similar to the antigenic advance~(\ref{DeltaT}) defined in the context of ferret data. With an additional measurement of the backbone titer $T^n_\alpha$, the set of mutant titers  $T^n_\beta = T^n_\alpha - \Delta T^n_{\alpha \beta}$ determines cross-immunities $c^n_\beta$ and, by equation (\ref{C_human}), a sample-averaged immunity profile $\bar C_\beta$. See also Note~4.

 \comment{
\subsubsection*{Human immune classes} 
The expressions (\ref{C_ferret}) and (\ref{C_human}) for the population immunity profile become equivalent if we partition humans into immune classes representing distinct, suitably coarse-grained infection and vaccination histories~\cite{Meijers2023}.} In the context of predictive analysis, class-based inference of immunity profiles has been applied to human antigenic data for SARS-CoV-2~\cite{Meijers2023}. In this case, individual samples are assigned a likely recent infection or vaccination dominating cross-immunity; the sample average (\ref{C_human}) can then be evaluated as an average over immune classes, equation~(\ref{C_ferret}), with weight functions informed by incidence tracking. More generally, the sample average (\ref{C_human}) may still give a {\em bona fide} input for fitness models; however, detecting and correcting biases in the population sample requires at least partial knowledge of the underlying infection histories by available metadata or computational reconstruction. This practical problem indicates a fundamental challenge of human immune data: given the complexity of human immunization histories~\cite{fonville2014antibody,lee2019mapping,dugan2020preexisting,auladell2022influenza}, only the most common patterns (e.g., the last vaccination or infection) will be represented correctly in realistic population samples. This limits the complexity of immune imprinting and antigenic sin that can be included in data-driven predictive analysis. \comment{Immune classes address this coarse-graining problem at a tunable level of complexity~\cite{Meijers2023}. In practice, we partition human antigenic data into classes} labelled by one or few previous immunization events, each associated with a specific viral clade, depending on the availability of metadata (Note~3).

\subsection{Evolutionary prediction and validation}

Predictive analysis harvests information from a system's past evolution to determine its likely near-future trajectory. At the core of predictions is a fitness model: a computable relation between independently measurable genetic and molecular traits as input and fitness, quantifying the expected relative growth of competing viral clades, as output. In this section, we first define absolute and relative fitness. Next, we describe a method to build consistent fitness models from distinct components, representing different sources of input data. Fitness models, together with tracking of the past dynamics, determine likely future frequency trajectories, as well as antigenic properties of emerging high-fitness variants. Finally, we discuss probabilistic measures to validate predictive methods and to quantify the time horizon of predictions.

\subsubsection*{Absolute and relative fitness}
The absolute fitness of a viral clade is defined as the growth rate of the number of infections, $F_\alpha (t) = \dot I_\alpha (t) / I_\alpha (t)$. By defining fitness at the level of clades, we average over systematic growth differences between strains within the same clade. Absolute fitness depends on the effective reproductive number, $R_\alpha (t)$ (the average number of new infections generated by an infected individual), and on the distribution of generational intervals (the time between infection and transmission)~\cite{park2019practical,park2022importance}. In the relevant parameter regime, this dependence is well approximated by $F_\alpha (t) = \log R_\alpha (t) /\tau$, where $\tau$ is the mean generational time interval~\cite{Meijers2023}. 

The relative fitness of a clade is defined with respect to the mean fitness of the viral population, 
\EQ
f_\alpha (t) = F_\alpha(t) - \sum_\beta x^\beta(t) F_\beta(t). 
\label{frel} 
\EE
Given that absolute fitness is proportional to the log of the reproductive numbers, relative fitness is invariant under a uniform rescaling $R_i \to a(t) R_i$, as expected for the temporal variation of reproductive numbers in the seasonal epidemics of influenza. Relative fitness governs the change of clade frequencies by selection, 
\EQ
\frac{\dot x^\alpha (t)}{x^\alpha (t)} = f_\alpha (t) + \dots; 
\label{xdot} 
\EE
here we have omitted stochastic changes by genetic drift \comment{and correction terms accounting for the spatial heterogeneity of absolute growth rates~\cite{Meijers2023}. } For sizeable clades of influenza, genetic drift can be neglected, and we obtain a deterministic relation between frequency trajectories and relative fitness that is the central equation of motion used for predictions. Given time-dependent frequency trajectories, we can also compute the change of the mean absolute fitness by evolution, $\Phi (t) = \sum_\alpha \dot x^\alpha (t) F_\alpha (t) = \sum_\alpha \dot x^\alpha (t) f_\alpha (t)$, the so-called fitness flux~\cite{mustonen2010fitness}. In the deterministic regime, the fitness flux equals the fitness variance in the population of circulating strains,
$\Phi (t) = {\rm Var} \, F(t)$, which is Fisher's fundamental theorem.

\subsubsection*{Fitness models}

In a fitness model, we write the absolute fitness of a clade as a sum of components that represent different molecular traits relevant for growth and can be quantified by independent input data. Here we use a simple two-component form 
\EQ
F_\alpha(t) = F_\alpha^\ant (t) + F_\alpha^\intr.
\label{Fdecomp} 
\EE
The antigenic fitness component, $F_\alpha^{\ant}(t)$, is generated by the cross-immunity built up in the population by previous infections and vaccinations and is, therefore, explicitly time- and history-dependent. The intrinsic fitness component, $F_\alpha^{\intr}$, collects the effects of other molecular traits that are relevant at specific points of the infection cycle. For example, changes in protein stability or host receptor binding affect the intra-host replication of the virus. Such changes are, in general, time-independent. In some of the literature, only the intrinsic fitness component is referred to as fitness, which is somewhat misleading.  The total fitness, not any single of its components, enters Fisher's theorem and governs the dynamics of clade frequencies. Specifically, as detailed below, the relative fitness (\ref{frel}) computed from the fitness model (\ref{Fdecomp}) can be compared to the empirical relative fitness (\ref{xdot2}) inferred from tracked frequency trajectories. 

\subsubsection*{Antigenic fitness} 
At the core of any predictive model is antigenic fitness, which has been a long-term driver of evolution in all human influenza lineages (and has recently become dominant in SARS-CoV-2 as well~\cite{Meijers2023}). Here we use the explicit inference of population immunity described above to evaluate this fitness component: the clade-specific antigenic fitness cost of the viral population is  proportional to the immunity profile $\bar C_\alpha (t)$ of the host population, 
\EQ
F_\alpha^\ant (t) = 
-\gamma_\ant \sum_\kappa  c_{\alpha}^\kappa \, r_\kappa(t), 
\label{Falpha} 
\EE
where $c_\alpha^\kappa = H (T_\alpha^\kappa)$ is given by equation~(\ref{H}). This form of the antigenic fitness contains the breakdown of $\bar C_\alpha (t)$ by immune classes, as given by equation~(\ref{C_ferret}). It reflects the clade-specific reduction of susceptible hosts and can be derived~\cite{Luksza2014, Meijers2023} from an underlying multi-strain epidemiological model~\cite{gog2002dynamics}. The antigenic fitness model is grounded upon the biophysics of molecular host-pathogen interactions, and it inherits the ``thermodynamic'' nonlinearity in the relation between titers and cross-immunity (section 3.3). The proportionality factor $\gamma_\ant$ sets the speed of evolution and is a free model parameter. 

Given antigenic data from a human population sample, $T_\alpha^n$ ($n = 1, \dots, N$), we can also evaluate the biophysical model of antigenic fitness from individual-based cross-immunity profiles sampled around a given time $t$, 
\EQ
F_{\alpha}^{\rm ag} (t) = - \frac{\gamma_{\rm ag}}{N} \sum_{n=1}^N c^n_\alpha , 
\label{F_human}
\EE
using the sigmoid map (\ref{cnalpha}) and the approximation of the population immunity profile $\bar C_\alpha (t)$ as a sample average, equation~(\ref{C_human}). This approach has first been used to infer viral fitness for SARS-CoV-2, again breaking down the data into likely immune classes (see Note~3)~\cite{Meijers2023}. For influenza, the correlation of  $\bar C_\alpha (t)$ with subsequent clade growth has been shown in ref.~\cite{kim2024}. 

Alternatively, antigenic fitness can be estimated from the HA protein sequence. Specifically, we can write the antigenic titer drop between a reference clade $\kappa$ and a viral clade $\alpha$ as a sum of mutational effects in antigenic epitopes, 
$\Delta T_{\alpha \kappa} = \sum_{m \in M^{\rm ep}_{\alpha, \kappa}} \varepsilon_m$, where $M^{\rm ep}_{\alpha, \kappa}$ is the set of epitope mutations separating clades $\alpha$ and $\kappa$; this additive form has been corroborated in~ref.~\cite{neher2016}. Assuming uniform mutational effects, antigenic fitness retains the biophysical form (\ref{Falpha}) with a sequence-based approximation of the cross-immunity matrix, $c_\alpha^\kappa = H (D^{\rm ep}_{\alpha, \kappa})$, 
where $D^{\rm ep}_{\alpha, \kappa}$ is the amino acid distance in HA epitope sequence between clades $\alpha$ and $\kappa$. In this approximation, equation~(\ref{Falpha}) reduces to the antigenic fitness model introduced in ref.~\cite{Luksza2014}. 

Genome-based models with position-specific effects can be informed by DMS data. In recent work, this method has been used to map immune escape from human antisera~\cite{welsh2023age,dadonaite2023full} and from monoclonal antibodies isolated from convalescent individuals~\cite{cao2023imprinted}. Inferring neutralization titers $T^n_{\alpha}$ from the escape scores of individual mutations~\cite{yu2022biophysical}, DMS data can be fed into the biophysical model~(\ref{F_human}) of antigenic fitness~\cite{Meijers2023,Kleist2023sars}. In other recent work, DMS escape scores have been directly related to empirical fitness differences between circulating clades~\cite{dadonaite2023full}. This implies a linear sequence-fitness map, $F_\beta^\ant \sim \sum_{k} \varepsilon (a_{k, \alpha}, a_{k, \beta})$, where $\alpha$ is the clade of the backbone strain and $\beta$ labels clades containing escape mutations.  
A common problem of additive sequence-based approximations is that epistatic effects between specific escape mutations present in a circulating strain are not systematically taken into account. See also Note~4.

\subsubsection*{Intrinsic fitness} 
Intrinsic fitness effects are a second component of fitness models. In sequence-based estimates, we can write the mutational load accumulated by HA protein changes again as a sum of mutational effects, 
\EQ
F_\alpha^\intr (t) = \sum_{k} s_k^\intr (a_{\alpha, k}). 
\EE
In the simplest approximation, we count all recent amino acid changes outside antigenic epitopes with uniform deleterious effects, $F_\alpha^\intr = \gamma_\intr D^{\rm ne}_{\alpha, A(\alpha)}$, where $A(\alpha)$ is a recent ancestor sequence of clade $\alpha$~\cite{Luksza2014}. Again, this estimate of mutational load can be improved by position- and amino-acid-specific effects measured in a DMS experiment~\cite{lee2018deep}. In a recent deep learning approach, antigenic and intrinsic fitness effects have been jointly inferred solely from HA protein sequence data~\cite{thadani2023learning}. See also Note~5.

\subsubsection*{Model calibration} 
In a second step, we calibrate the model-based fitness by comparison with empirical trajectory data $x^\alpha_r (t)$ in a representative set of regions $r$ where clades are in direct competition (see section 3.1). We define a new, calibrated fitness model $F^c$ by 
\EQ
F^c_{\alpha,r} (t) = F_{\alpha, r} (t) + \Delta F_\alpha.
\label{cal}
\EE
The calibration combines clade- and region-specific trajectories of the primary model, here the two-component model $F_{\alpha, r} (t) = F_{\alpha, r}^\ant (t) + F_\alpha^\intr$ computed with region-specific immune weights, and clade-specific  global correction terms $\Delta F_\alpha$. \comment{The calibrated model inherits the explicit time-dependence of the antigenic fitness component, while the correction terms are taken to be time-independent.} Relative fitness computed from the calibrated model, $f^c_{\alpha,r} (t)$, is then used as input for predictions.

The calibration algorithm operates by minimizing an error score over a suitable period up to the prediction baseline, 
\EQ
S = \int^{t_0} \Big [ \sum_{\alpha, r} (\dot x^\alpha_r (t) - f^c_{\alpha, r} (t) x^\alpha_r (t) ) Y_r (t) \Big ]^2  {\rm d}t, 
\label{Scal}
\EE
where $f^c_{\alpha, r} (t) = F^c_{\alpha,r} (t) - \sum_\beta F^c_{\beta,r} (t) \, x^\beta_r (t)$ and $Y_r(t)=N_r(t) / \sum_{r'} N_{r'} (t)$ is the fraction of sequences collected from region $r$ at time $t$. We use a suitable prior to constrain non-zero corrections to few, significant values.

The calibration of model-based and empirical fitness serves three purposes: 
(1) The correction term systematically screens the set of trajectories for significant fitness effects that are not contained in the explicit fitness model. The specific  calibration scheme given in equation (\ref{cal}) identifies global, time-independent mismatches of model-based and empirical fitness; examples are occasional positive fitness effects caused by intrinsic changes. Another example of potential mismatches are antigenic effects of neuraminidase changes, which are not mapped by HI assays. 
(2) The calibration yields optimal values of model parameters, here $\gamma_\ant$ and $\gamma_\intr$.  
(3) A variant of this method without primary model input and without prior can be used as a standalone inference method of empirical fitness from tracking data. This provides global, model-free fitness estimates $\hat f_\alpha (t)$ from regional frequency trajectories. The data partitioning by region is key for correct output, because the inference of relative fitness from global trajectories is confounded by differences in absolute growth between regions~\cite{Meijers2023}.

\subsubsection*{Evolutionary predictions}
Clade frequency predictions require frequency tracking, $x^\alpha (t)$, and a fitness model $F_\alpha (t)$ informed by data up to a prediction baseline at time $t_0$ (we suppress the index $c$ for calibrated models). The corresponding relative fitness serves to predict the most likely future frequency trajectories,
\EQ
x^\alpha(t) = x^\alpha(t_0) \exp \left[ \int_{t_0}^{t} f_\alpha(t') dt' \right], 
\label{pred} 
\EE
over a limited time interval into the future of the prediction baseline. This relation follows by integration of equation~(\ref{xdot}) in the deterministic limit. Frequency tracking gives the initial condition $x^\alpha(t_0)$, the fitness model determines the future changes of the predicted trajectories. 

The precise genetic changes that spawn new variants have a stochastic origin and may difficult to predict. The antigenic characteristics of successful emerging variants are more predictable, because the population immunity profile constrains of antigenic evolution. Specifically, as shown in ref. \cite{Meijers2023}, temporal windows of strong antigenic selection in a given immune class $\kappa$ is generated when high population immunity coincides with high expected loss of cross-immunity on the steep flank of the landscape $H(T)$ given by equation~(\ref{H}). Therefore, tracking of population immunity trajectories $C^\kappa_\alpha(t)$ can give important information on the direction of the next antigenic escape mutations. 

A dedicated web platform, Previr (\href{https://previr.app}{https://previr.app}), reports continuously updated tracking and and fitness predictions obtained by the computational pipeline described here for human influenza A(H3N2), A(H1N1)pdm09, B(Vic), as well as SARS-CoV-2. See Note~6.

\subsubsection*{Model validation}
Given a set of historical predictions, each based on data collected before a specific baseline $t_0$, we can compare the predicted trajectories with their posterior tracked counterparts. In this way, we can test prediction methods and quantify the so-called {\em prediction horizon} $\tau$; that is, the characteristic temporal range of reliable predictions~\cite{Luksza2014,lassig2017predicting,lassig2023steering}. Care is to be taken that predictions are solely computed from data in the period up to $t_0$; for example, a strain tree computed from input and validation data would violate this requirement by propagating posterior information back into the training period. 

We can quantify the quality of prediction by the {\em predictive information}, a probabilistic measure of how much a given model reduces the uncertainty about future trajectories~\cite{Luksza2014,lassig2017predicting,lassig2023steering}. Importantly, this measure separates the information obtained by tracking, which constrains the starting point of future evolutionary paths at time $t_0$, from the contribution of the fitness model, which explains a part of the evolutionary change from $t_0$ to $t_0 +\tau$. It also serves to define the prediction horizon as the characteristic time scale on which predictions show diminishing information return. 

The prediction horizon of current predictions is limited due to the emergence of new variants. Historical predictions based on a two-component fitness model of the form~(\ref{Fdecomp}) have a prediction time span of order 1 year~\cite{Luksza2014, lassig2017predicting}. Alternatively, the empirical fitness inferred from recent growth data before the prediction baseline, $\hat f_\alpha (t_0)$, can be used as a standalone method for short-term predictions \cite{neher2014predicting}. This method is independent of the specific assumptions of a fitness model, but it does not capture the non-linearities and time-dependence of antigenic fitness~\cite{barrat2024eco}. Recent method developments on predicting antigenic characteristics of new mutations can potentially increase the prediction horizon.

\subsection{Vaccine protection}
Evolutionary predictions as described above can be combined with antigenic data to select the vaccine composition for influenza, and in a similar way for SARS-CoV-2. Predictive modelling is  essential for this decision, because the vaccine strain needs to pre-empt the most likely future viral population. Here we derive measures of vaccine protection against present and future viral populations that serve to rank candidate vaccines. These measures can be computed from tracked or predicted trajectories of viral evolution and population immunity.

\subsubsection*{Naive protection profiles}
In analogy to the cross-protection by primary infections (section 3.3), we define the (cross-)protection of a vaccine as the relative reduction in the susceptibility to infections after vaccination. Resolving protection at the level of clades, we compute the protection profile of a given vaccine from antigenic data, 
 \EQ
 c_\alpha^\vac = H(T_\alpha^\vac),
 \label{cvac}
 \EE
using the nonlinear map~(\ref{H}). This profile applies to a population of otherwise naive hosts, because we single out vaccination as the only source of primary immunization. Given the protection profile of a vaccine, we can also define the mean protection against circulating strains at a given time, 
\EQ
\bar c^{\rm vac}(t) = \sum_\alpha x^\alpha(t) c_\alpha^{\rm vac}.
\label{eta} 
\EE
Importantly, this measure is time-dependent because clade frequencies evolve over time.  The cross-immunity factors $c_\alpha^\vac$ measure the maximum protection of a given vaccine against strains of a given clade. This value is reached for infections shortly after vaccination, before waning of vaccine-generated humoral immunity sets in. Immune waning generates a time-dependent protection profile $c_\alpha^{\rm vac} (t)$, introducing an additional time-dependence in the mean protection $\bar c^{\rm vac}(t)$ \comment{(Note~7). }

Figure 5A shows the protection profile $c_\alpha^\vac$ computed by equation~(\ref{cvac}) for two successive vaccine components for influenza A(H3N2). The vaccine component based on the reference strain A/Cambodia/e0826360/2020 has been recommended for the Northern Hemisphere (NH) 21/22 season, the updated vaccine component based on the strain A/Darwin/6/2021 for the Southern Hemisphere (SH) 22 and 23 seasons and the NH 22/23 and 23/24 seasons \cite{whorecommendation} (these strains are marked by syringes). The protection profile of the A/Cambodia/e0826360/2020-like strain shows the partial escape of later circulating strains from vaccine protection, the updated A/Darwin/6/2021-like vaccine restores this protection. In Fig.~5B, we show another example, the protection profiles of the successive vaccine components A/Hawaii/70/2019 (NH 20/21 season) and A/Wisconsin/588/2019 (SH 21 and 22 seasons, NH 21/22 and 22/23 seasons) for influenza A(H1N1)pdm9. These vaccine strains have complementary protection profiles: the A/Hawaii/70/2019-like strain protects against the strains predominant before 2021, the A/Wisconsin/588/2019-like strain against later strains. In both examples, the corresponding mean protection trajectories, $\bar c^{\rm vac}(t)$, of the successive vaccine components show an opposite time dependence (Fig.~5C,D). All of these protection estimates are computed retrospectively and are based on ferret data, which measure vaccine protection without pre-existing immunity. 

\begin{figure}[ht!]
\centering
\includegraphics[width=.75\linewidth]{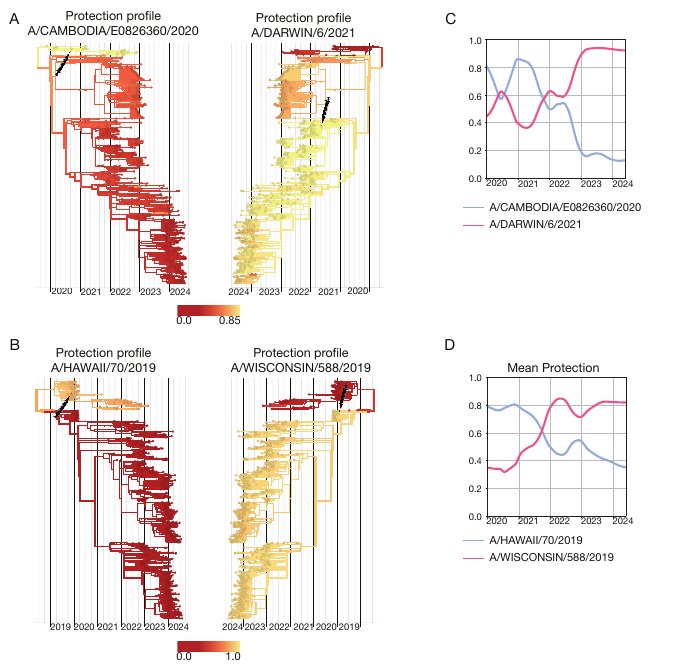}
\caption{\textbf{Cross-protection of successive vaccine strains.} 
(A) Protection profiles, $c^\vac_\alpha$, of the influenza A(H3N2) A/Cambodia/e0826360/2020-like (left) and A/Darwin/6/2021-like (right) vaccine components, mapped onto the strain tree. Shading from yellow to red indicates decreasing cross-protection; vaccine strains are marked on the tree by syringe symbols.
(B) A(H1N1)pdm09 A/Hawaii/70/2019-like (left) and A/Wisconsin/588/2019-like (right) vaccine components. 
(C,D) Corresponding trajectories of mean vaccine protection, $\bar c^{\rm vac}(t)$. Based on ferret antigenic data for influenza A(H3N2) from ref.~\cite{WIC}.
} 
\end{figure}

\begin{figure}[ht!]
\centering
\vspace*{-2mm}
\includegraphics[width=.9\linewidth]{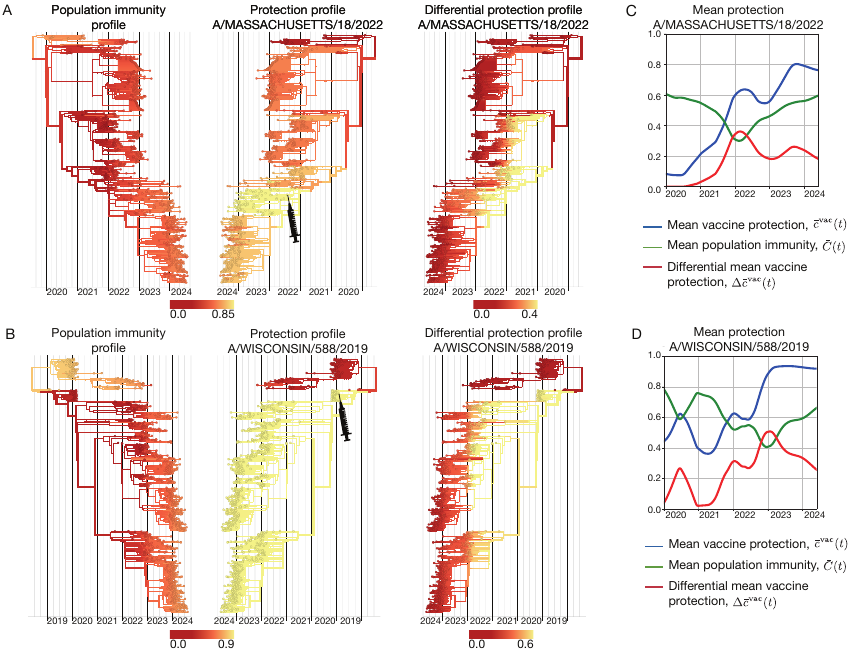}
\vspace*{5mm}
\caption{\textbf{Differential protection of vaccines.}  
(A)~Influenza A(H3N2) population immunity profile, $\bar C_\alpha (t)$ (left); protection profile $c^\vac_\alpha$ of the A/Massachusetts/18/2022-like vaccine component (center); resulting differential vaccine protection profile $\Delta c^\vac_\alpha (t)$ (right).
(B)~Influenza A(H1N1)pdm09 population immunity profile, $\bar C_\alpha (t)$ (left); protection profile $c^\vac_\alpha$ of the A/Wisconsin/588/2019-like vaccine component (center); resulting differential vaccine protection profile $\Delta c^\vac_\alpha (t)$ (right).
(C,D)~Trajectories of mean population immunity, $\bar C(t)$ (green), mean vaccine protection, $\bar c^{\rm vac}(t)$ (blue) and differential mean vaccine protection, $\Delta \bar c^{\rm vac}(t)$ (red). Based on ferret antigenic data for influenza A(H3N2) from ref.~\cite{WIC}.
}
\end{figure}

\subsubsection*{Differential protection profiles}
We can also consider the protection profile of a given vaccine, defined again as the relative reduction in the susceptibility to $\alpha$-infections after vaccination, in a population with pre-existing immunity. This measure, the differential protection profile $\Delta c^\vac_\alpha (t)$, is computed as 
\EQ
\Delta c^\vac_\alpha (t) = 
\sum_\kappa r_\kappa (t) \max [c_\alpha^{\rm vac} - c_{\alpha}^\kappa, 0 ].  
\label{Delta_c_alpha}
\EE
Here the cross-immunity factors $c_{\alpha}^\kappa$ and the immune class weights $r_k (t)$ determine pre-existing immunity as in equation~(\ref{C_ferret}), and  $c_\alpha^{\rm vac}$ is the naive vaccine protection given by equation~(\ref{cvac}). We assume that in a given individual, the immuno-dominant repertoire component activated in response to an infecting strain $\alpha$ is induced by the vaccination or by previous immunizations, whichever component generates the stronger cross-immunity. \comment{This immunodominance assumption gives an approximate form for a polyclonal B cell immune response shaped by immune imprinting. It follows from a recent biophysical model of antibody lineage activation in an acute infection that predicts an affinity-graded activation of multiple lineages, generating a correlation between antigen affinity and clone size~\cite{moran2023immune}. The resulting differential protection profile is a host population average of immunodominance, }$\Delta c^\vac_\alpha (t) = \frac{1}{N} \sum_{n =1}^N  \max [c_\alpha^{\rm vac} - c_\alpha^n, 0]$; the form (\ref{Delta_c_alpha}) then follows by grouping individuals into immune classes. We can also evaluate the differential mean protection against circulating strains, 
\EQ
\Delta \bar c^{\rm vac}(t) = \sum_\alpha x^\alpha(t) \, \Delta c^\vac_\alpha (t),
\label{Delta_eta} 
\EE 
which quantifies the added value of vaccination over population immunity. This measure is time-dependent for two reasons: population immune dynamics affect $\bar C_{\alpha}(t)$ and viral evolution propagates the frequencies $x^\alpha (t)$. Notably, a good vaccine is expected to decrease in differential protection over time, as population immunity builds up against viral clades $\alpha$ covered by the vaccine. 

By relating the differential mean protection to the total pre-vaccination population immunity $\bar C (t) = \sum_\alpha x^\alpha(t) \, \bar C_\alpha (t)$, we obtain a computational estimate of vaccine effectiveness (VE) \cite{Jackson2013,Trombetta2022} in the population: the expected case numbers of vaccinated and unvaccinated individuals are proportional to $1 - \bar C (t) - \Delta \bar c^{\rm vac}(t)$ and $1 - \bar C(t)$, respectively, which gives 
\EQ
{\rm VE } \, (t) = \frac{\Delta \bar c^{\rm vac}(t)}{ 1 - \bar C (t)}.
\label{VE}
\EE 
Importantly, vaccine effectiveness is seen to be a dynamical quantity: it can increase when the naive protection $\bar {c}^{\rm vac}(t)$ increases and decrease when population immunity against clades covered by the vaccine builds up. In a similar way, we can compute vaccine efficacy as measured in clinical trials, by calibrating the mean protection $\bar C (t)$ to known characteristics of the trial cohort.  \comment{To include the waning of vaccine-induced immunity, we evaluate equations~(\ref{Delta_c_alpha}) and~(\ref{Delta_eta}) with a time-dependent vaccine protection profile $c_\alpha^{\rm vac} (t)$ (Note~7). These predictive VE estimates can be computed from molecular and surveillance data available before a given influenza season. They do not take into account external factors such as timing and severity the seasonal epidemic. Relative vaccine effectiveness, here differences between candidate vaccines based on different vaccine strains, is expected to be less sensitive to these external factors than the absolute effectiveness of a given vaccine. Therefore, the main use of predictive VE estimates is the pre-emptive ranking of candidate vaccines, as described in the next subsection. 
}

In Fig.~6A, we compare the protection profile $c^\vac_\alpha$ of the A/Massachusetts/18/2022-like vaccine component (SH 24 season, NH 24/25 season) with the underlying population immunity profile $\bar C_{\alpha}(t)$ of influenza A(H3N2). These profiles result in the differential protection profile $\Delta c_\alpha^{\rm vac}(t)$, as given by equation~(\ref{Delta_c_alpha}). In this case, the population immunity profile is broader but weaker than vaccine-induced protection, which generates added protection peaked around the vaccine clade (marked by a syringe). The corresponding analysis of the A/Wisconsin/588/2019-like vaccine component for influenza A(H1N1)pdm09 is shown in Fig.~6B. In the season 2021-22, population immunity and vaccine induced immunity cover complementary viral clades; the vaccine adds protection specifically against clades subject to weak population immunity. We can also track the corresponding trajectories of mean population immunity, $\bar C(t) = \sum_\alpha x^\alpha (t) \bar C_{\alpha}(t)$, and vaccine protection, $\bar c^{\rm vac}(t)$ and $\Delta \bar c^{\rm vac}(t)$ (Fig.~6C,D). In both examples, the differential vaccine protection shows a diminishing added value over time, in parallel to the buildup of population immunity against newer viral clades.

\subsubsection*{Pre-emptive ranking of candidate vaccines}
The mean naive and differential protection, $\bar c^{\rm vac} (t)$ and $\Delta \bar c^{\rm vac}(t)$, are directly related to viral fitness. By equations~(\ref{Falpha}) and~(\ref{eta}), the mean naive protection $\bar c^{\rm vac}(t)$ is proportional to the vaccine-induced drop of the mean viral population fitness in a naive human population. The mean differential protection $\Delta \bar c^{\rm vac}(t)$, equation~(\ref{Delta_eta}), measures the analogous drop of viral fitness in a population with pre-existing immunity. Hence, $\Delta \bar c^{\rm vac}(t)$ is an explicitly co-evolutionary measure, which depends on the distribution of circulating viral strains and on the distribution of human immunity. Here we have established a general method to compute estimates of vaccine protection for a rapidly changing population of viral strains and population immunity states. The specific evaluation of these measures is likely to undergo changes, as new types of input data become available. 

The vaccine protection measures $\bar c^{\rm vac} (t)$ and $\Delta \bar c^{\rm vac}(t)$ can be used to rank available candidate vaccines for a given influenza season. By equation (\ref{VE}), the ranking by  $\Delta \bar c^{\rm vac}(t)$ is equivalent to a ranking by VE. Importantly, vaccine protection changes over time due to viral evolution and population immunity dynamics, underscoring the role of predictive analysis for vaccine strain selection. By using predicted clade frequencies, $x^\alpha(t)$, and predicted immune profiles, $\bar C_\alpha (t)$, we can optimize vaccines in a pre-emptive way, taking into account the likely viral evolution and population immune dynamics up to and throughout the season when the vaccine will be used. In addition, we can account for the partial waning of vaccine protection between vaccination and the expected next exposure (Note~7).
The protection profiles shown in Fig.~5 and~6 provide a more detailed picture than the mean protection, highlighting viral clades with maximal escape from vaccine-induced and pre-existing population immunity.

\section{Notes}

\begin{enumerate}

\item 
Due to intense sequencing efforts during the past decade, over 100,000 influenza and several million SARS-CoV-2 genomes have become available to the scientific community. Keeping track of the genetic variation observed in sequenced viral samples is a bioinformatic bottleneck~\cite{Hodcroft2021}; the challenges for computational analysis will increase with time. Two recently published methods, MAPLE~\cite{DeMaio2023} and USHER~\cite{Turakhia2021}, overcome the bioinformatic bottleneck using efficient algorithmic solutions within a maximum-parsimony framework. These approaches do not include the inference of ancestral sequences and the timing of internal nodes, which represents a second computationally expensive process. To address the high computational demand of maximum-likelihood approaches, a common procedure is to subsample available data to a few thousand sequences. This method is adequate for displaying major clades of the evolving viral population, but subsampling may miss clades at small population frequency, including recently emerged, fast-growing clades. Here we implement a maximum-likelihood inference procedure for a subtree partitioning of the global tree, which avoids the need for subsampling.

\item Glossary of antigenicity measures. 
(1) Cross-neutralization matrix, $T_{\alpha}^\kappa$: average neutralization titers of antisera from immune class $\kappa$ against viral strains from clade $\alpha$. 
(2)~Cross-immunity matrix, $c_\alpha^\kappa$: expected cross-protection, or reduction of susceptibility, of an individual from immune class $\kappa$ against viral strains from clade $\alpha$.
(3)~Population immunity trajectories, $C_\alpha^\kappa(t)$: population immunity, or reduction of viral growth, for strains of clade $\alpha$ induced by previous infections by strains from clade $\kappa$. 
(4)~Population immunity profile, $\bar C_{\alpha}(t)$: total population immunity against strains from clade $\alpha$. 
(5)~Average population immunity, $\bar C (t)$: net population immunity against the population of circulating strains.
(6)~Vaccine protection profile, $c_{\alpha}^\vac$: cross-protection, or reduction of susceptibility, against viral strains from clade $\alpha$ induced by a given vaccine (vac) in a naive host. Results in a time-dependent mean protection, $\bar c^{\rm vac} (t)$, against circulating strains. 
(7)~Differential vaccine protection profile, $\Delta c_{\alpha}^\vac(t)$: cross-protection, or reduction of susceptibility, against viral strains from clade $\alpha$ induced by a given vaccine (vac) in a population with pre-existing immunity profile $\bar C_{\alpha}(t)$. Results in a time-dependent differential mean protection, $\Delta \bar c^{\rm vac} (t)$, against circulating strains. 

\item The established animal model for influenza uses ferrets that are immune-naive against influenza prior to the infection with a reference strain. Neutralization tests of these ferret antisera capture vaccination-induced cross-protection against test strains without confounding effects of pre-existing immunity. The use of ferret data in fitness models for viral evolution in humans depends on a heuristic extrapolation: cross-immunity $c^\kappa_\alpha$ computed from ferrets infected with a strain from clade $\kappa$ is a good proxi for humans who had their last immunization (infection or vaccination) by a strain from the same clade. More complex immunization histories are, for example, combinations of a recent infection and vaccination; such cases can be modelled by composite immune classes, $\kappa = (\kappa_1, \kappa_2)$~\cite{Meijers2023}. An important role of the last immunization in the response to future exposures has been established~\cite{fonville2014antibody}. Older exposures may also be relevant in specific population segments~\cite{zhang2019original,welsh2023age}.  The impact of immune complexity on viral dynamics has recently been studied in ref.~\cite{barrat2024eco}.  

\item Deep mutational scanning (DMS) \cite{starr2022deep, cao2023imprinted} and high-throughput in-vitro evolution \cite{moulana2023genotype,zahradnik2021sars} are emerging as promising data sources for predictions of viral evolution. These data map the genomic distribution of high-fitness mutations and can inform antigenic and intrinsic components of genome-based fitness models. 
Using specific additivity assumptions, antigenic escape scores of individual point mutations can be processed to fitness estimates for strains with multiple mutations, including variants that have not yet been observed as circulating strains or have not yet been phenotypically characterized. However, this construction of fitness landscapes neglects epistasis between specific escape mutations. Other challenges in applying DMS data to predictive analysis of influenza evolution include the choice of the backbone strain from which the mutant library is constructed, as well as the choice of representative antisera driving the escape dynamics. 

\item Key intrinsic fitness components, including protein stability and binding to human receptors, depend on protein structure. Computational models hold promise for a high-throughput evaluation of structure-based phenotypic and fitness effects~\cite{greenbaum2008,greenbaum2014,harvey2023bayesian,Armita2024,tsai2024glycan}. A recent model combines computational estimates of such effects with deep learning to predict the genetic evolution of influenza and SARS-CoV-2 \cite{thadani2023learning}. Machine-learning tools to predict protein structures, like AlphaFold \cite{Alphafold2021}, can also inform predictions \cite{akdel2022structural}, but a successful application of these methods to evolutionary analysis requires sufficient accuracy in distinguishing wild-type and mutated protein structures~\cite{mcbride2023alphafold2, buel2022can, pak2023using}.

\item The effective application of this prediction pipeline depends on comprehensive surveillance of influenza sequence evolution, antigenic evolution, and epidemiology. The collaborative efforts of researchers, national public health institutions, and the global WHO surveillance and response system are key to maintain global surveillance under comparable standards. The timely availability of sequence and antigenic data is key to flag new variants of concern early in their trajectory, or even before their appearance \cite{Meijers2023}. Surveillance is also necessary to support DMS and laboratory evolution experiments, to allow updating of backbone strains and of reference antisera generating immune pressure, in tune with the evolving viral population.

\item \comment{Vaccine-induced humoral immunity reaches a maximum shortly after vaccination, which is followed by a decay process called immune waning. The form and decay rate of immune waning will depend on the immunization type and pathogen. For SARS-CoV-2, we can describe waning as an exponential process with rate $\nu$, corresponding to a linear reduction of log titers, $T_\alpha^\vac (t, t_{\rm vac}) = T_\alpha^\vac - \nu (t - t_{\rm vac})$ for $t > t_{\rm vac}$~\cite{Meijers2023}. This form has been comprehensively recorded~\cite{iyer2020persistence,israel2021large}. 
Waning generates a time-dependent protection profile, here 
$c_\alpha^{\rm vac} (t) = \int^t r (t_{\rm vac}) \, H(T_\alpha^\vac (t, t_{\rm vac})) \, d t_{\rm vac} / \int^t r (t_{\rm vac}) \, d t_{\rm vac}$, which replaces the form (\ref{cvac}) and introduces an additional time-dependence in the protection measures (\ref{eta}) -- (\ref{VE}). Here $r (t_{\rm vac})$ denotes distribution of vaccination times in a given season. For predictive analysis, we estimate this distribution from an average over past seasons.  }
\end{enumerate}

\subsection*{Acknowledgements}
\small{We thank all personnel at the Worldwide Influenza Centre, WHO Collaborating Centre for Reference and Research on Influenza, situated at The Francis Crick Institute, involved in the generation and sharing of antigenic data. Special thanks go to Nicola Lewis, Ruth Harvey, Monica Galiano, and John McCauley for discussions on multiple aspects of this work. We also thank all National Influenza Centers (NICs) and institutions who have contributed information, clinical specimens and viruses, epidemiological, and other associated data to the WHO Global Influenza Surveillance and Response System (GISRS), including FluNet. We are grateful to all institutions and individuals who have contributed clinical specimens, virus sequences, and associated data through the GISAID EpiFlu database. 

This work was funded in part by Deutsche Forschungsgemeinschaft grant CRC 1310, Collaborative Research Center {\em Predictability in Evolution} (to ML), the Centers of Excellence for Influenza Research and Response (CEIRR) contract \#75N93021C00014 (to MŁ and ML), the National Institute of Allergy and Infectious Diseases, National Institutes of Health (NIAID, NIH) grant R01AI165820 (to MŁ). MŁ was funded in part as a Pew Biomedical Scholar.}

\end{document}